\documentclass[twocolumn,english,aps,prl,showpacs,floatfix, preprintnumbers,amsmath,amssymb]{revtex4-1}
\usepackage{mathrsfs}
\usepackage{stmaryrd}
\usepackage[dvips]{graphicx} 
\usepackage{subfigure}
\usepackage{dcolumn}
\usepackage{bm} 
\usepackage{babel}

\DeclareMathAlphabet{\mathpzc}{OT1}{pzc}{mb}{it}
\begin{document}

\title{Coherence freeze in an optical lattice investigated via pump-probe spectroscopy}
\author{Samansa Maneshi}
\email[]{smaneshi@physics.utoronto.ca}
\author{Chao Zhuang}
\author{Christopher R. Paul}
\author{Luciano S. Cruz}
\altaffiliation[Current address: ]{UFABC, S\~ao Paulo, Brazil.}
\author{Aephraim M. Steinberg}
\affiliation{Centre for Quantum Information \& Quantum Control  and  Institute for Optical Sciences,  \\Department of Physics, University of Toronto, Canada }

\date{\today}
\begin{abstract}
Motivated by our observation of fast echo decay and a surprising coherence freeze, we have developed a   pump-probe spectroscopy technique for vibrational states of ultracold $^{85}$Rb atoms in an optical lattice to gain information about the memory dynamics of the system. We use pump-probe spectroscopy to monitor the time-dependent changes of frequencies experienced by atoms and to characterize the probability distribution of these frequency trajectories. We show that the inferred distribution, unlike a naive microscopic model of the lattice, correctly predicts the main features of the observed echo decay. 
\end{abstract}
\maketitle 
Characterizing decoherence mechanisms is a crucial task for experiments aiming to control quantum systems, e.g., for quantum information processing (QIP).  In this work, we demonstrate how two-dimensional (2D) pump-probe spectroscopy may be extended to provide important information on these mechanisms.  As a model system, we study quantum vibrational states of ultracold atoms in an optical lattice.  In addition to being a leading candidate system for QIP \cite{BrennenJaksch}, optical lattices are proving a versatile testing ground for the development of quantum measurement and control techniques \cite{OMandel, Anderlini} and a powerful tool for quantum simulations, e.g. the study of Anderson localization and the Hubbard model \cite{MottAnderson}.  \par
In our experiment, we study the vibrational coherence of  $^{85}$Rb atoms trapped in a shallow one-dimensional standing wave. Through our 2D pump-probe technique, we obtain detailed microscopic information on the frequency drift experienced by atoms in the lattice, enabling us to predict the evolution of coherence. Since the pioneering development of the technique in NMR\cite{Jeener-Ernst}, 2D spectroscopy has been widely used 
to obtain high-resolution spectra and gain information about relaxations, couplings, and many-body interactions, in realms ranging from NMR \cite{Ernst} to molecular spectroscopy \cite{Mukamel-Jonas, Hybl, Brixner, MillerNature} to semiconductor quantum wells \cite{Cundiff, KWStone}. Here, we show that similar powerful techniques can be applied to the quantized center-of-mass motion of trapped atoms, and more generally, offer a new tool for the characterization of systems in QIP and quantum control. 
\par
\begin{figure}[b]
\includegraphics[scale=0.275]{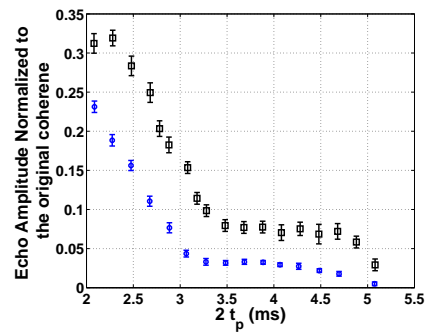}
\caption{(Color online) Two typical measurements of echo amplitude vs. time. The echo pulse and the observed echo envelope are  centered at times $t_p$ and $2t_p$, respectively. After an initial decay, echo amplitude stays constant for about $1ms$ forming a plateau, before decaying to zero. The average lattice depths are $20E_R$ (circles) and $18E_R$ (squares).} 
\label{fig1}
\end{figure}
We have previously measured the evolution of coherence between the lowest two vibrational states of potential wells \cite{Ours}. 
The dephasing time is about $0.3ms$  ($T^{\star}_2$). 
This dephasing is partly due to an inhomogeneous distribution of lattice depths as a result of the transverse Gaussian profile of the laser beams. To measure the homogeneous decoherence time ($T_2$), we perform pulse echoes, measuring the echo amplitude as a function of time  \cite{Ours}.
Figure~\ref{fig1} shows two typical measurements of echo amplitude carried out on different dates under slightly different conditions such as different average lattice depths and different dephasing times. The echo amplitude  initially decays with a time constant of about $0.7ms$, which is much faster than the photon scattering time ($\sim 60ms$) in the lattice. It then exhibits a $1ms$-long coherence freeze followed by a final decay. 
Absent real decoherence on the short time scale of $1ms$, only loss of frequency memory would inhibit the appearance of echoes. This loss comes about when atoms experience  time-varying frequencies. We use 2D pump-probe spectroscopy to monitor this frequency drift. Our 2D pump-probe spectroscopy is essentially a version of spectral hole-burning for vibrational states. By monitoring the changes in the hole spectrum as a function of time we gain information on the atoms' frequency drift.
Information obtained from our 2D spectra enables us to characterize the temporal decay of frequency memory and through our simulations we find that ``coherence freeze" is related to the shape of this memory loss function. 
Similar plateaus in echo decay and a two-stage decay of echo amplitude have been observed in a Cooper-pair box \cite{Nakamura}, for a single electron spin in a quantum dot \cite{Vandersypen} and for electron spins in a semiconductor \cite{SClark}. Those plateaus or two-stage decays have been either explained through {\it{a priori}} models or simply described phenomenologically. Here, we are introducing an experimental technique to directly probe the origin of plateaus.  

The periodic potential in our experiment is formed by interfering two laser beams blue-detuned by $ 25GHz$ from the D2 transition line, $F=3 \shortrightarrow F^{\prime}=4$ ($\lambda=780nm$), thus trapping atoms in the regions of low intensity, which minimizes the photon scattering rate and the transverse forces. The two laser beams intersect with parallel linear polarizations at an angle of $\theta = (49.0 \pm 0.2)^{\circ}$, resulting in a spacing of $L=(0.930 \pm 0.004) \mu m$ between the wells. Due to gravity, the full effective potential also possesses a ``tilt'' of $2.86 E_R$ per lattice site, where $E_R=\frac{h^2}{8mL^2}$ is the effective lattice recoil energy. The photon scattering time in our experiment is  $\approx 60ms$ and the Landau-Zenner tunneling times for transitions from the lowest two levels are greater than $ 160ms$. 
Atoms are loaded to the lattice during a molasses cooling stage and prepared in the ground vibrational state by adiabatic filtering \cite{StefanQPT}. Due to the short coherence length of atoms in optical molasses ($60 nm$ at $10 \mu K$), there is no coherence between the wells. 
We measure populations of atoms in the ground vibrational, the first excited, and the (lossy) higher excited states $P_1$, $P_2$, and $P_{L}$, respectively, by fluorescence imaging of the atomic cloud after adiabatic filtering \cite{StefanQPT}. \par

The pump and probe pulses are sinusoidal phase modulations of one of the laser beams forming the lattice. The modulation is of the form $\phi(t)=A(t)[1-cos(\omega_m t)]$, where $A(t)$ is a square envelope function with amplitude $2\pi/72$ and $\omega_m$ is a variable frequency. The duration of each pulse is $8$ cycles, i.e., $T=8 (2\pi/\omega_m)$. This phase modulation shakes the lattice back and forth periodically, coupling vibrational states of opposite parity. To first-order in modulation amplitude, the phase-modulating part of the Hamiltonian  has the same form as the electric dipole Hamiltonian. 

\begin{figure}
\includegraphics[scale=0.25]{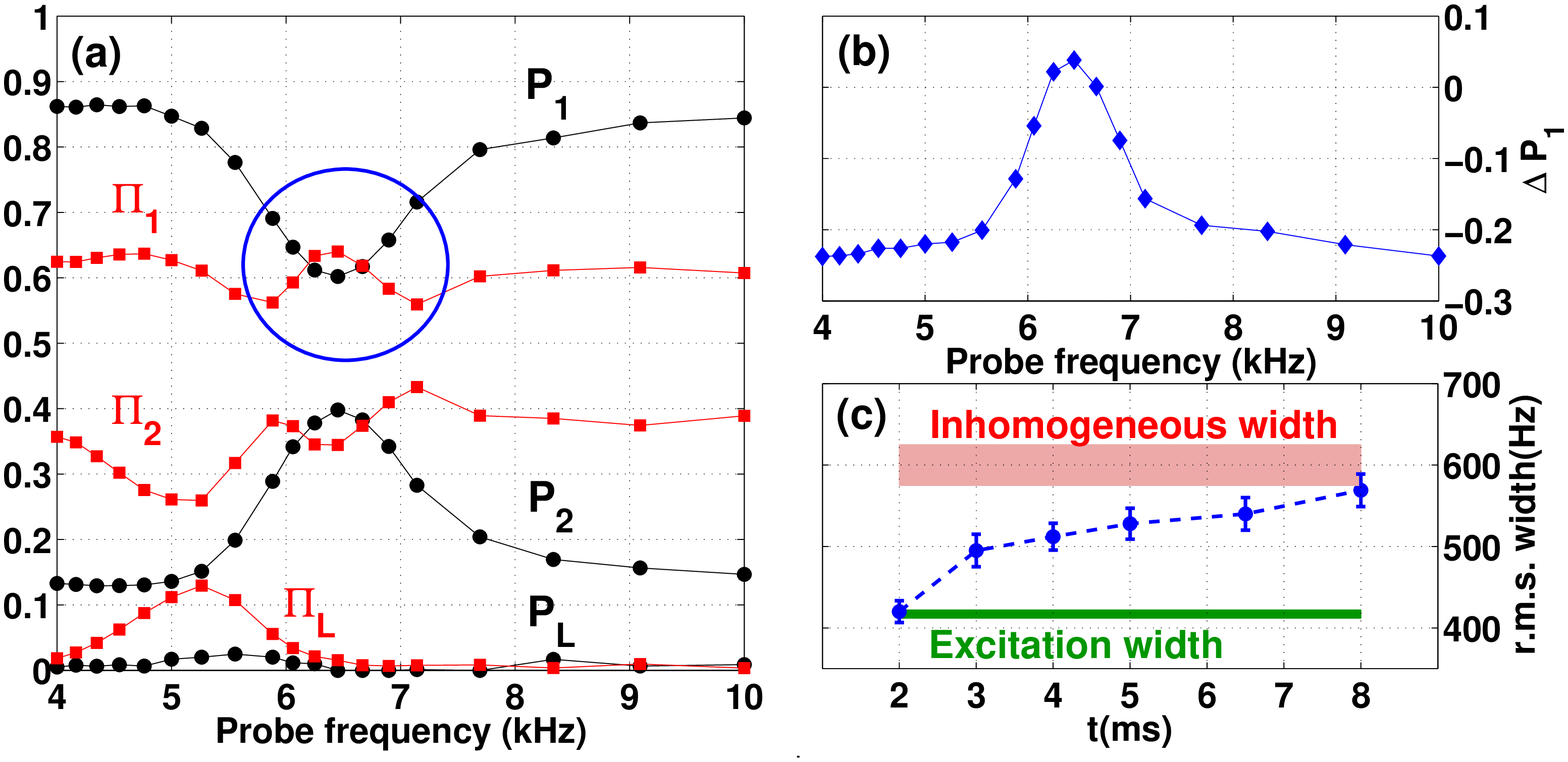}
\caption{(Color online) Experiment:(a) black circles: probe-alone populations, $P_1, P_2,$ and  $P_{L}$. Red squares: pump-probe populations, $\Pi_1, \Pi_2, \Pi_{L}$ for a pump at $6.45kHz$ and a delay of $2ms$. The circled region shows the hole-burning signal in $P_1$ and $\Pi_1$. (b) The difference spectrum $\Delta P_1 = \Pi_1 -P_1$. (c) Growth in the r.m.s. width of a Gaussian fit to $\Delta P_1$ as a function of delay. The bar centered at $(600 \pm 25)Hz$ is the inhomogeneous width determined from the $P_1$ spectrum. The green line centered at $(417 \pm 5)Hz$ indicates the instrumental width expected due to the spectra of our pulses.}
\label{fig2}
\end{figure}

\begin{figure}[b]
\includegraphics[scale=0.4]{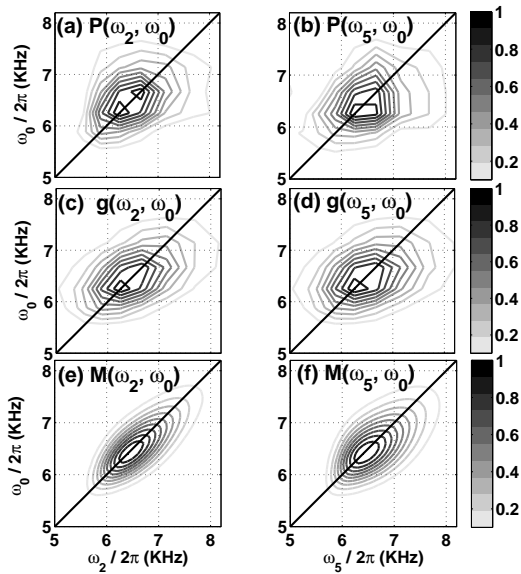}
\caption{Contour plots: measured joint probabilities (a) $P(\omega_2, \omega_0)$, (b) $P(\omega_5, \omega_0)$; convolved marginals (c) $g(\omega_2, \omega_0)$, (d) $g(\omega_5, \omega_0)$ obtained by convolving bare marginals $M(\omega_n, \omega_0)$ with the power spectra of the pump and probe pulses; bare marginals (e) $M(\omega_2, \omega_0)$, (f) $M(\omega_5, \omega_0)$.}
\label{fig3}
\end{figure}

The inhomogeneous spectrum of vibrational excitations is measured in an average lattice depth of $24E_R$ by applying probe pulses at different frequencies and measuring state populations. Figure~\ref{fig2}(a) shows state populations $P_1$, $P_2$, and $P_{L}$ (black circles) as a function of probe frequency.  
We then measure the pump-probe spectrum for a fixed delay. 
A pump pulse with a specific frequency is applied, exciting atoms in wells whose vibrational transition frequency matches the frequency of the pump pulse, therefore burning a hole in the spectrum of the ground state population. 
After a delay, probe pulses at different frequencies are applied, coupling the ground and excited states of atoms whose frequencies match those of the probe pulses. 
The red squares in Fig.~\ref{fig2}(a) show populations $\Pi_1$, $\Pi_2$, and $\Pi_{L}$ as a function of probe frequency  for a delay of $2ms$ and at a pump frequency of $6.45kHz$. 
The central part of $\Pi_1$ shows the hole burnt into the spectrum of the ground state population. 
To characterize the hole, we plot the difference between the pump-probe and the probe-alone spectra, $\Delta P_1= \Pi_1-P_1$, in Fig.~\ref{fig2}(b). We monitor the frequency drift of atoms by changing the delay between the pump and probe pulses. Figure~\ref{fig2}(c) shows the r.m.s. width of $\Delta P_1$ as a function of delay. The r.m.s. width increases with increasing delay until it approaches the inhomogeneous width of the lattice. For pump-probe delays shorter than $2ms$, the coherence present between the lowest two states results in Ramsey fringes making it impractical to extract useful spectra from the measurements. 
Repeating the above procedure for different pump frequencies, we obtain a 2D spectrum of $\Delta P_1$ for a fixed delay as a function of $\omega_0$, the pump frequency at time $t_0$, and $\omega_n$, the probe frequency at $t_n$. 
We interpret the measured 2D spectrum as the conditional probability, $P(\omega_n|\omega_0)$, of an atom having frequency $\omega_n$ at $t_n$ given that at time $t_0$ its frequency was $\omega_0$. The joint probability is $P(\omega_n, \omega_0)=P(\omega_n|\omega_0) P(\omega_0)$, where $P(\omega_0)$ is the probe-alone spectrum. 
Figures~\ref{fig3}(a-b) show the experimental joint probabilities $P(\omega_2, \omega_0)$ and $P(\omega_5, \omega_0)$, for delays of $2ms$ and $5ms$, respectively. By measuring joint probabilities for three different delays, $P(\omega_2, \omega_0)$, $P(\omega_3, \omega_0)$ , and $P(\omega_5, \omega_0)$, we construct a joint distribution of three frequencies at three different times  in order to characterize how 
frequencies change as a function of time (frequency trajectories) in the first $5ms$. Details of our analysis are provided in the next two  paragraphs.
 \par
%
We assume a skew-normal distribution  $\mathpzc{SN}(\boldsymbol{\omega})$ \cite{Azzalini}, which is a generalization of the normal distribution. The multivariate skew-normal distribution is defined as $\mathpzc{SN}(\boldsymbol{\omega})=2 \psi(\boldsymbol{\omega}; \pmb{\Omega}) \Psi(\boldsymbol{\alpha}^T \boldsymbol{\omega})$, where  $\psi$ is the three-dimensional normal density function, and $\Psi$ is the cumulative distribution function. $\boldsymbol{\omega}=(\boldsymbol{\tilde{\omega}}- \boldsymbol{\mu})$, where  $\boldsymbol{\tilde{\omega}}=(\omega_0, \omega_2, \omega_5)$ denotes frequencies at three different times.  $\boldsymbol{\mu}=(\mu_0, \mu_2, \mu_5)$ and $\boldsymbol{\alpha}=(\alpha_0, \alpha_2, \alpha_5)$ are the three frequency means and shape parameters, respectively.
$\pmb{\Omega}$ is the covariance matrix whose diagonal terms are variances $\{\sigma_{i}^2\}$, and whose off-diagonal terms are covariances  $\{\sigma_{i}\sigma_{j} \rho_{ij}\}$, where $\{\rho_{ij}\}$ are the mutual correlation coefficients. We estimate the best values for these parameters by using a genetic algorithm, as described below. \par
The bivariate marginals of $\mathpzc{SN}(\boldsymbol{\omega})$, denoted as $M(\omega_n, \omega_m)$, are calculated by integrating $\mathpzc{SN}$ with respect to the third frequency, e.g., $M(\omega_2, \omega_0)=\int d\omega_5 \, \,  \mathpzc{SN}(\boldsymbol{\omega})$. 
These marginals are then convolved with the power spectra of the pump and probe pulses to obtain the three bivariate convolved marginals $g(\omega_n, \omega_m)$ for comparison to the three measured spectra, $P(\omega_n, \omega_m)$. Similarly, the univariate marginals of $\mathpzc{SN}(\boldsymbol{\omega})$ are calculated and convolved with the power spectra of the probe pulses for comparison to the three measured probe-alone spectra,  $P(\omega_n)$.  
Then, the sum of the squared residuals for all data points  is calculated. The convolved marginal $g(\omega_5, \omega_2)$ is compared to the experimental $P(\omega_3, \omega_0)$ with the assumption that the frequency changes experienced by atoms in the $(2  \shortrightarrow  5)ms$ interval are the same as in the $(0  \shortrightarrow  3)ms$ interval. 
From repeated runs of the genetic algorithm function in MatLab, we find $12$ sets of best fit parameters, and take  their average.  
Figures~\ref{fig3}(c-f) show 
the two convolved marginals $g(\omega_2, \omega_0)$,  $g(\omega_5, \omega_0)$ and the corresponding bare marginals $M(\omega_2, \omega_0)$, $M(\omega_5, \omega_0)$ thus obtained. 
The distribution parameters are means $\{\bar{\mu}_0, \bar{\mu}_2, \bar{\mu}_5\}=\{5.92(4), 5.92(3), 5.96(6)\} kHz$; r.m.s. widths $\{\bar{\sigma}_0, \bar{\sigma}_2, \bar{\sigma}_5\}=\{0.77(5), 0.82(4), 0.83(7)\} kHz$; mutual correlation coefficients $\{\bar{\rho}_{02}, \bar{\rho}_{25}, \bar{\rho}_{05}\}= \{0.88(2), 0.82(3), 0.80(4)\}$; and shape parameters $\{\bar{\alpha}_0, \bar{\alpha}_2, \bar{\alpha}_5\}= \{2.6(5), 3.2(4), 4.0(2.0)\}$. The numbers in parentheses are the standard deviations calculated from the $12$ optimization runs.
The means and r.m.s. widths stay approximately constant. Furthermore, the mutual correlations $\bar{\rho}_{25}$ and $\bar{\rho}_{05}$ are essentially equal and within two standard deviations of $\bar{\rho}_{02}$. The increase in the anti-diagonal width of $M(\omega_5, \omega_0)$ compared to $M(\omega_2, \omega_0)$ is a signature of the slight decrease in the mutual correlation coefficient $\bar{\rho}_{05}$ compared to $\bar{\rho}_{02}$. 
\par
\begin{figure}[b]
\includegraphics[scale=0.375]{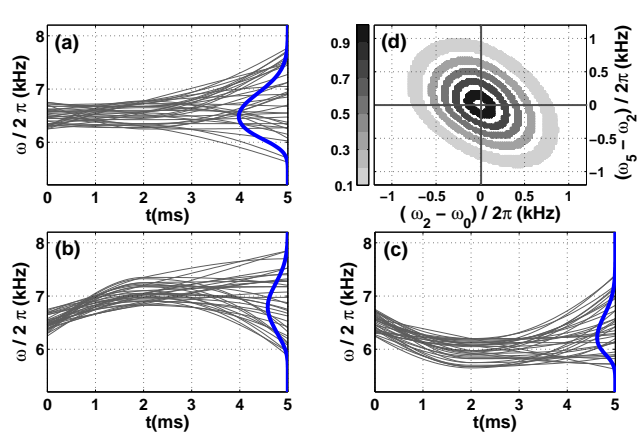}
\caption{(Color online) (a,b,c) (ccw from top left): Samples of smoothed frequency trajectories selected according to their probabilities. Trajectories are constrained to originate from the frequency range $\Delta \omega_0=0.5kHz$ and to pass through a window of width $\Delta \omega_2=0.52kHz$ at $t=2ms$.  $\Delta \omega_0$ is centered at $6.5kHz$ and $\Delta \omega_2$ is centered at: (a) $6.5kHz$, (b) $7.07kHz$, and (c) $5.93kHz$.
The sideways curves show the final probabilities $P_f(\omega_5)$.
(d) Projection of $\mathpzc{SN}(\boldsymbol{\omega})$ onto the plane defined by frequency differences $( \omega_{2} - \omega_0)$ and $(\omega_{5} - \omega_2)$, showing a negative correlation between the two time intervals.}
\label{fig4}
\end{figure}  
The bare joint distribution $\mathpzc{SN}(\boldsymbol{\omega})$ is our estimate of the probability of frequency trajectories. 
Since $\mathpzc{SN}$ is a function of frequencies at three different times, frequency trajectories are constructed by interpolating between frequencies at these times for all possible combinations of $\omega_0$, $\omega_2$, and $\omega_5$. 
Figures~\ref{fig4}(a-c) show three groups of smoothed \cite{pchip} trajectories that are constrained to originate from the frequency range $\Delta \omega_0$ and to pass through a window of width $\Delta \omega_2$ at $t=2ms$. Drawn for each group are small samples of trajectories selected according to their probability. The sideways curves are the final probabilities at $t=5ms$ calculated from $P_f(\omega_5) = \int_{\Delta \omega_0} d \omega_0 \int_{\Delta \omega_2} d\omega_2 \, \mathpzc{SN}(\boldsymbol{\omega})$. 
The most probable trajectories (Fig.~\ref{fig4}(a)) stay within one standard deviation (width of the univariate skew-normal marginal at $t=0$) from the center of distribution. The next most probable trajectories (Fig.~\ref{fig4}(b)) are the ones where frequency initially increases in the first $2ms$ followed by trajectories where frequency decreases in the first $2ms$ (Fig.~\ref{fig4}(c)). 
Figure~\ref{fig4}(d) is a projection of  $\mathpzc{SN}(\boldsymbol{\omega})$ onto the plane defined by frequency differences,  $(\omega_2- \omega_0)$ and $(\omega_5- \omega_2)$. Each quadrant corresponds to either an increase or decrease in the two time intervals, $(0 \shortrightarrow 2)ms$ and $(2 \shortrightarrow 5)ms$. The negative correlation between the two time intervals indicates that a rising frequency in the first time interval is usually followed by a falling frequency in the second interval and vice versa. This is in sharp contrast to the positive correlation we would obtain by considering the ballistic expansion of atoms through a spatial Gaussian distribution of well depths. In that case, there are no trajectories where an initial frequency  decrease is followed by an increase. Although ballistic expansion would give rise to similar time scales for echo amplitude decay it does not predict the formation of plateaus. This shows the power of 2D spectroscopy in characterizing all frequency trajectories without reference to any {\it a priori} model. \par
\begin{figure}[t]
\includegraphics[scale=0.4]{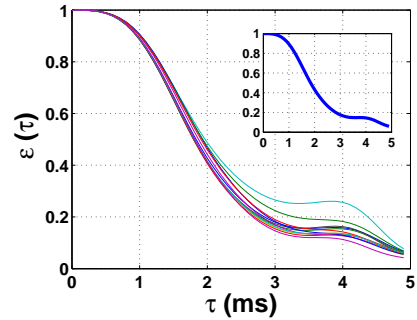}
\caption{(Color online) A selection of echo curves, $\epsilon (\tau)$, generated by different $\mathpzc{SN}(\boldsymbol{\omega})$ resulting from repeated optimization trials. The inset shows the echo decay curve generated by $\mathpzc{SN}(\boldsymbol{\omega})$ calculated from the average values of parameters.}
\label{fig5}
\end{figure}

The echo amplitude can be predicted by calculating the net accumulated phase, $\varphi( \tau\, ;\, \boldsymbol{\tilde{\omega}})$, along each trajectory before and after the echo pulse, and by weighting each trajectory according to its probability $\mathpzc{SN}(\boldsymbol{\omega})$,
\begin{equation}
\epsilon(\tau) = \left| \int d\boldsymbol{\tilde{\omega}}  \, \,  \mathpzc{SN}(\boldsymbol{\omega}) \, \, e^{-i\varphi(\tau \, ;\, \boldsymbol{\tilde{\omega}})} \right|
\label{echoamp}
\end{equation}
where $\varphi( \tau\, ;\, \boldsymbol{\tilde{\omega}})=\int_{t_1}^{\tau =2t_1} dt'' \omega(t'') -\int_0^{t_1} dt' \omega(t') $. 
As Fig.~\ref{fig5} shows, the decay time constant of echo amplitude predicted by the bare probability distribution $\mathpzc{SN}(\boldsymbol{\omega})$ is similar to the measured value. 
Depending on the exact method of interpolation (smooth or linear), the details of the echo decay vary but we always see a two-stage decay. We hypothesize that the non-decaying values of mutual correlation coefficients $\bar{\rho}_{25}$ and $\bar{\rho}_{05}$ lead to the formation of the plateau. By repeating simulations with reduced values of  $\bar{\rho}_{25}$ and $\bar{\rho}_{05}$, we find that for values of $(\bar{\rho}_{25} - \bar{\rho}_{05} > 0.4)$ the plateau disappears. 
A more accurate calculation of echo decay would require  knowledge of the joint probability distribution of multiple frequencies at multiple times on a finer time scale. The complexity of the task of obtaining data and extracting a joint distribution grows rapidly as the time scale is made finer. The extracted $\mathpzc{SN}(\boldsymbol{\omega})$ is our best estimate for a trivariate joint distribution. An exact reconstruction of the distribution requires knowledge  of higher-order conditional probabilities such as $P(\omega_5 | \omega_2, \omega_0)$. In principle, it would be possible to extract this conditional probability from a
two-pump, one-probe measurement. Simulations, however, show that this signal might be very difficult to extract in practice \cite{mythesis}. \par
%
In conclusion, we have developed a two-dimensional pump-probe spectroscopy technique for vibrational states of ultracold atoms in optical lattices to gain information about the memory dynamics of the system without any need for {\it{a priori}} physical models. Our method is a direct frequency-domain characterization of the system for different delays between the pump and probe pulses. By measuring populations of the quantum vibrational states,  we infer information about the underlying correlations,  which in turn enables us to predict how coherence evolves in the system. 
By directly extracting a best estimate trivariate joint distribution $\mathpzc{SN}(\boldsymbol{\omega})$ from the measured 2D spectra at three different delays, and constructing trajectories accordingly, we have calculated echo amplitude as a function of time. The inferred distribution correctly predicts the main features of the observed echo decay and it suggests that persistence of frequency-frequency correlations leads to the formation of plateaus. The similarity between the predicted echo decay curve and the directly measured ones demonstrates the power of 2D spectroscopy in providing information on memory dynamics, information which could not always have been foreseen. 
Our 2D pump-probe method is a general technique that should be broadly applicable in a variety of quantum information candidate systems for characterization of decoherence mechanisms.    \par  
%
We would like to thank Alex Paarmann for helpful discussions and we acknowledge financial support from NSERC, QuantumWorks, and CIFAR. LSC acknowledges support from CNPq, Brazil.

\end{document}